\documentclass[preprint]{jfm}

\usepackage{graphicx}
\usepackage{dcolumn}
\usepackage{bm}
\usepackage{latexsym}
\usepackage{subfloat}
\usepackage{subfig}
\usepackage{psfrag}
\usepackage{xcolor}
\usepackage{color}
\usepackage{natbib}
\usepackage{epstopdf}
\usepackage{amsmath}
\usepackage{wasysym}

\ifCUPmtlplainloaded \else
  \checkfont{eurm10}
  \iffontfound
    \IfFileExists{upmath.sty}
      {\typeout{^^JFound AMS Euler Roman fonts on the system,
                   using the 'upmath' package.^^J}%
       \usepackage{upmath}}
      {\typeout{^^JFound AMS Euler Roman fonts on the system, but you
                   dont seem to have the}%
       \typeout{'upmath' package installed. JFM.cls can take advantage
                 of these fonts,^^Jif you use 'upmath' package.^^J}%
      }
  \else
  \fi
\fi

\ifCUPmtlplainloaded \else
  \checkfont{msam10}
  \iffontfound
    \IfFileExists{amssymb.sty}
      {\typeout{^^JFound AMS Symbol fonts on the system, using the
                'amssymb' package.^^J}%
       \usepackage{amssymb}%
         \let\leq=\leqslant
         
      }{}
  \fi
\fi

\ifCUPmtlplainloaded \else
  \IfFileExists{amsbsy.sty}
    {\typeout{^^JFound the 'amsbsy' package on the system, using it.^^J}%
     \usepackage{amsbsy}}
    {}
\fi

\newsavebox{\astrutbox}
\sbox{\astrutbox}{\rule[-5pt]{0pt}{20pt}}

\title[Three-dimensional Lagrangian Vorono\"{\i} analysis for clustering]{Three-dimensional Lagrangian Vorono\"{\i} analysis for clustering of particles and bubbles in turbulence} 

\author[Y. Tagawa, J. Mart\'inez, V. N. Prakash, E. Calzavarini, C. Sun, and D. Lohse]%
{Y\ls O\ls S\ls H\ls I\ls Y\ls U\ls K\ls I\ns T\ls A\ls G\ls A\ls W\ls A$^{1,3}$\footnote{y.tagawa@utwente.nl},\ns  
J\ls U\ls L\ls I\ls \'A\ls N\ns  M\ls A\ls R\ls T\ls \'I\ls N\ls E\ls Z\ns M\ls E\ls R\ls C\ls A\ls D\ls O$^{1,3}$,\ns \break
V\ls I\ls V\ls E\ls K\ns  N.\ns P\ls R\ls A\ls K\ls A\ls S\ls H$^{1,3}$,\ns 
E\ls N\ls R\ls I\ls C\ls O\ns  C\ls A\ls L\ls Z\ls A\ls V\ls A\ls R\ls I\ls N\ls I$^{2,3}$,\ns \break 
C\ls H\ls A\ls O\ns S\ls U\ls N$^{1,3}$\footnote{c.sun@utwente.nl},\ns \and D\ls E\ls T\ls L\ls E\ls F\ns L\ls O\ls H\ls S\ls E$^{1,3}$\footnote{d.lohse@utwente.nl}\ns}
\affiliation{$^1$Physics of Fluids Group, Faculty of Science and Technology, J.M. Burgers Center for Fluid Dynamics, University of Twente, PO Box 217, 7500 AE  Enschede, The Netherlands\\[\affilskip]
$^2$ Laboratoire de M\'ecanique de Lille CNRS/UMR 8107, Universit\'e Lille 1  and  Polytech'Lille, Cit\'e Scientifique Av.~P.~Langevin, 59650 Villeneuve d'Ascq, France\\[\affilskip]
$^3$International Collaboration for Turbulence Research}
  
\date{\today}

\begin{document}
\maketitle
\begin{abstract}
Three-dimensional Vorono\"{\i} analysis is used to quantify the clustering of inertial particles in homogeneous isotropic turbulence using data sets from numerics in the point particle limit and one experimental data set. We study the clustering behavior at different density ratios, particle response times (i.e. Stokes numbers St) and two Taylor-Reynolds numbers ($Re_\lambda$ = 75 and 180).
The Probability Density Functions (PDFs) of the Vorono\"{\i} cell volumes of light and heavy particles show a different behavior from that of randomly distributed particles ---i.e. fluid tracers---implying that clustering is present. 
The standard deviation of the PDF normalized by that of randomly distributed particles is used to quantify the clustering. The clustering for both light and heavy particles is stronger for the higher $Re_{\lambda}$.
Light particles show maximum clustering for St around 1$-$2 for both Taylor-Reynolds numbers.
The experimental dataset shows reasonable agreement with the numerical results.
The results are consistent with previous investigations employing other approaches to quantify the clustering. 
We also present the joint PDFs of enstrophy and Vorono\"{\i} volumes and their Lagrangian autocorrelations. The small Vorono\"{\i} volumes of light particles correspond to regions of higher enstrophy than those of heavy particles, indicating that light particles cluster in higher vorticity regions. 
	The Lagrangian temporal 
autocorrelation function of Vorono\"{\i} volumes shows that
 the clustering of light particles lasts  much longer than that of  heavy or neutrally 
buoyant particles. Due to inertial effects arising from the density contrast with the surrounding liquid, 
light and heavy particles
remain clustered for much longer times than the flow structures which cause the clustering.

\end{abstract}

\section{Introduction}\label{sec:intro}
The distribution of particles transported by turbulent flows is a current research topic with implications in diverse fields, such as process technology (\cite{Pratsinis1996}), cloud formation (\cite{Bodenschatz2010}), and plankton dynamics (\cite{Schmitt2008}).  In most of the cases, the particles have a finite size and a different density than the carrier fluid, i.e. they have inertia. These inertial particles cannot totally follow the fluid motion and distribute inhomogeneously within the turbulent flow, leading to clustering or preferential concentration (\cite{Toschi2009a}). The two relevant dimensionless parameters describing the dispersed inertial particles in the fluid are the density ratio $\beta=3\rho_f/(\rho_f+2\rho_p)$, where $\rho_f$ and $\rho_p$ are the densities of the carrier fluid and particle, respectively, and the Stokes number, St$=\tau_p/\tau_{\eta}$, where $\tau_p=a^2/3\beta\nu$ is the particle relaxation time, $\tau_{\eta}$ is the typical timescale of the flow, which for a turbulent flow is the Kolmogorov time scale, $a$ is the particle radius, and $\nu$ is the kinematic viscosity of the fluid.

In recent years, both numerical and experimental studies have quantified the clustering of particles by employing different approaches like statistical analysis of single-point measurements (\cite{Calzavarini2008b}), box-counting method (\cite{Fessler1994,Aliseda2002}), pair correlation functions  (\cite{Chen2006,Saw2008}), Kaplan-Yorke dimension  (\cite{Bec2006,Calzavarini2008c}), Minkowski functionals (\cite{Calzavarini2008c}) and segregation indicators (\cite{Calzavarini2008a, IJzermans2009}). It is not possible to obtain global information on bubble clustering from a single-point analysis (\cite{Calzavarini2008b}).  Methods like box-counting and pair correlation functions, although useful, require the selection of an arbitrary length scale that affects the quantification of the clustering. The Kaplan-Yorke dimension, based on the calculation of the Lyapunov exponents, quantifies the contraction of a dynamical system by considering the separation rates of particle trajectories. Nevertheless, it does not provide global morphological information. Minkowski functionals, originally used to provide complete morphological information of the large-scale distribution of galaxies (\cite{Kerscher2001}), have been applied to study the clustering of particles in turbulent flows (\cite{Calzavarini2008c}). \cite{Calzavarini2008c} found that light particles cluster in filamentary structures, whereas heavy particles have a wall-like topology around interconnected tunnels, and obviously no clustering was observed for neutrally buoyant tracers. In the above numerical simulations and experiments, the strongest clustering was found for particles with St$\approx$O(1). The problem with Minkowski-type analysis is that it is numerically expensive, and it does not provide information on the Lagrangian evolution of the clusters.

An alternative mathematical tool that can be used to study clustering is the Vorono\"{\i} tessellation, which has been used in astronomy as a tool to characterize clustering of galaxies (\cite{Weygaert1989}). Recently, \cite{Monchaux2010} apply a Vorono\"{\i} analysis to quantify the clustering of heavy particles in grid-generated turbulence. This Vorono\"{\i} approach does not require the selection of an arbitrary length scale for a fixed particle number, and it can provide information on the Lagrangian statistics of clustering \cite[]{Monchaux2010}. \cite{Monchaux2010} obtain two-dimensional particle positions by imaging a turbulent flow in a wind tunnel seeded with droplets. The Vorono\"{\i} cells are defined based on the positions of the particles within the measurement domain. One can quantify the clustering by calculating the probability density function (PDF) of the normalized areas of the Vorono\"{\i} cells. The PDF will have a different shape for inertial particles when compared to the corresponding PDF of randomly distributed particles. The main difference is observed at the small and large values of normalized areas, where the PDF of heavy particles has a higher probability than for randomly distributed particles. There is a central region where there is no significant difference between the PDFs of heavy particles and randomly distributed ones. The values of normalized areas at which the PDF deviates from the randomly distributed particles can be used as thresholds to classify Vorono\"{\i} cells that belong either to clusters or voids. \cite{Monchaux2010} report a maximum preferential concentration for St around unity, in agreement with other methods that have been used to study clustering. 

The objective of the present work is to extend the work of \cite{Monchaux2010} to: (i) three-dimensions and (ii) a much larger range of density ratios (including light, heavy, and neutrally buoyant particles) and Stokes numbers, i.e. we quantify particle clustering by applying 3D Vorono\"{\i} analysis both for numerical and experimental data sets of particles and bubbles. Moreover, we (iii) correlate the clustering behavior of different particles with local turbulent flow quantities and (iv) study the Lagrangian temporal evolution of the clusters.

\section{Experimental and Numerical Datasets and Vorono\"{\i} analysis} \label{sec:exp}
\subsection{Datasets}

\begin{table}
\caption{Summary of the simulation and experimental parameters, where $N$ is the size of the numerical domain, $Re_{\lambda}$ is the Taylor-Reynolds number, $\eta$, $\tau_{\eta}$ are the Kolmogorov length and time scales, respectively, and $ N_{particles}$ is the number of particles in the simulations, and the time-averaged particle number in the measurement volume for the experiment.}
\begin{center}
 \begin{tabular}{c|c|c|c|c|c|c|c}
 \hline 
                     &$N$    &$Re_{\lambda}$&$\eta$	&$\tau_{\eta}$				&$ N_{particles}$	&$St $\\ \hline 
         Simulation A  & 128 &   75  			& 0.0332 	& 0.1104 						&$1.0\times10^3 $ 	&0.1 - 4.0	\\
         Simulation B  & 512 & 180 			& 0.001   	& 0.0483  						&  6.4$\times10^4$	&0.1 - 4.1\\
Experiment &  -     & 162 			&     288 $\mu m$       	          	&	80 ms				&   1.3$\times10^3$	& 0.04 $\pm$ 0.02\\ \hline 
    \end{tabular}
    \end{center} \label{tab:flowCond}
\end{table}

The numerical scheme for a dilute suspension (neglecting particle collisions) of point particles in homogeneous and isotropic turbulence is described as follows (\cite{Maxey1983,Calzavarini2008c}):

\begin{equation}
\label{particle model}
\frac{d\mbox{\boldmath $v$}}{dt}=\beta\frac{D}{Dt}\mbox{\boldmath$u$}(\mbox{\boldmath$x$}(t),t) -\frac{1}{\tau_p}(\mbox{\boldmath $v$}-\mbox{\boldmath $u$}(\mbox{\boldmath$x$}(t),t))
\end{equation}
where $\mbox{\boldmath$v$}=d\mbox{\boldmath$x$}/dt$ is the particle velocity and $\mbox{\boldmath$u$}(\mbox{\boldmath$x$}(t),t)$ the velocity field.
The dimensionless numbers used to model the particle motion are the density difference between the particle and the fluid $\beta$ and the Stokes number St. The values of $\beta$ = 0, 1 and 3 correspond to very heavy particles, neutrally buoyant tracers, and bubbles in water, respectively. When St = 0, the particles perfectly follow the fluid flow behaving as fluid tracers. As summarized in table~\ref{tab:flowCond}, we explore a parameter space of $\beta$ = 0, 1, and 3 and St ranging from 0.1 to 4.0 consisting of 24 values at $Re_{\lambda}$ = 75 with the spatial resolution of $N =128^3$. For $Re_{\lambda}$ = 180 with $N =512^3$, we study 5 different values of St = 0.1, 0.6, 1.6, 2.6, and 4.1.
\cite[from iCFDdatabase http://cfd.cineca.it,][]{Calzavarini2008c}. The simulation of the Navier-Stokes equation is based on a 2/3 de-aliased pseudo-spectral algorithm with  2$^{nd}$ order Adams--Bashforth time-stepping (for details see  \cite{Bec2006}).  Simulations have been performed in a cubic box of side $L = 2\pi$ with periodic  boundary conditions.
The forcing adopted acts only at the largest scale, it is implemented by keeping constant the kinetic energy content in of the smallest shell ($|k|$$\leq$ 1) in Fourier space. 
The intensity of the forcing is adjusted in such a way to have a turbulent dissipative scale ($\eta$) of about 0.8 lattice grids in real space.
Particle dynamics is evolved with time steps $O(10)$  times smaller than the smallest Stokes time, leading to an accurate resolution of the particle trajectories.
Tri-linear interpolation is used to determine the value of the velocity field at the particle position.  
The numerical code was also validated by comparison against an independent code implementing different temporal integration scheme, different particles interpolation and different large scale forcing (\cite{Toschi2009b}). The simulations extends over few O(1) large-eddy-turnover times, this is enough for particles  to reach a statistically steady distribution. 
In the present analysis, we fix the number of particles ($N_{particles}$) for given Reynolds numbers: 1000 particles for the simulation with the domain size of 128$^3$ at $Re_\lambda$ = 75, and 6.4$\times$10$^4$ particles for the simulation of 512$^3$ at $Re_\lambda$ = 180. The number of particles normalized by the corresponding domain volume, i.e. the volume concentrations of the particles, for the two $Re_\lambda$ are identical. In one particular case of $Re_{\lambda}$ =75 and St = 0.6, the particle number is varied from 100 to 1 $\times$ 10$^5$.

We conduct experiments in  the Twente Water Tunnel (TWT), an 8 m long vertical water tunnel designed for studying two-phase flows. By means of an active grid, nearly homogeneous and isotropic turbulence  with $\mathrm{Re}_{\lambda}$ up to 300 can be achieved. A  measurement section  with dimensions 2$\times$0.45$\times$0.45 m$^3$ with three glass walls provides optical access for the three-dimensional particle tracking velocimetry (PTV) system.  Micro-bubbles  with a mean radius of 170 $\pm 60$ $\mathrm{\mu}$m are generated by blowing pressurized air through a ceramic porous plate that is located  in the upper part of the water tunnel. These micro-bubbles are advected downwards by the flow passing through the measurement section.
In our 3D-PTV micro-bubble experiments, we use a 4-camera system to get micro-bubble positions in the active-grid-generated turbulence in the TWT. The experimental data are collected for a duration of 6 seconds (three times the large eddy turnover time) at an acquisition rate of 1,000 fps.
For the experimental data, $Re_{\lambda}$ = 162, $\beta$ = 3 and St = 0.04 $\pm$ 0.02, and the time-averaged number of particles inside the measurement volume of 70mm$\times$70mm$\times$70mm is $1.3\times10^3$ (for further details, see \cite{Martinez2010, Martinez2011}).

\subsection{Vorono\"{\i} analysis}
\begin{figure}
\begin{center}
\includegraphics[width=70mm]{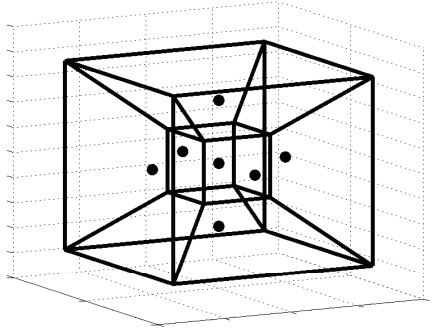}
\end{center}
\caption{ An example of a 3D Vorono\"{\i} tesselation. The dots represent the particle position and lines represent the borders of the Vorono\"{\i} cells.} 
\label{3DVoronoi}
\end{figure}

The Vorono\"{\i} diagram is a spatial tesselation where each Vorono\"{\i} cell is defined at the particle location based on the distance to the neighboring particles (\cite{SpatialTesselation}). 
Every point in a Vorono\"{\i} cell is closest to the particle position compared to the neighboring particles, the exceptions are the vertices, borderlines and facets (see figure~\ref{3DVoronoi}). Therefore, in regions where particles cluster, the volume of the Vorono\"{\i} cells is smaller compared to that of the cells in neighboring regions. Hence, the volume of the Vorono\"{\i} cells is inversely proportional to the local particle concentration. 
The PDF of the Vorono\"{\i} volumes normalized by the mean volume for randomly distributed particles 
can be well described by a $\Gamma$-distribution (\cite{Ferenc2007}) (see figure~\ref{betaDependency4}). In the 3D case,  the $\Gamma$-distribution has the following prefactor and exponent:
\begin{equation}
\label{Gamma function}
f(x)=\frac{3125}{24}x^4 \exp (-5x).
\end{equation}
Here $x$ is the Vorono\"{\i} volume normalized by the mean volume. Particles which are not randomly distributed will have a PDF that deviates from this $\Gamma$-distribution, indicating preferential concentration. The Vorono\"{\i} cells of particles located near the edges of the domain are ill-defined, i.e. they either do not close or close at points outside of the domain. These cells at the border of the domain are not considered for the analysis. 
 
\section{Results}

{First, we present results on the effect of the density ratio ($\beta$) on the clustering, followed by the effect of the Stokes number (St) and the number of particles ($N_{particles}$). Then, we show how the volume of Vorono\"{\i} cells ($\mathcal{V}$) and enstrophy are related. Finally, we present results on the Lagrangian autocorrelations of Vorono\"{\i} volumes and enstrophy.

\subsection{Density effect}
Here we study the clustering behavior of particles of different $\beta$ at  a fixed St for two different $Re_{\lambda}$. Figure~\ref{betaDependency4} shows the PDFs of the Vorono\"{\i} volumes ($\mathcal{V}$) normalized by their averaged volume ($\mathcal{\bar{V}}$), $\mathcal{V}$/$\mathcal{\bar{V}}$, for heavy, neutrally buoyant, and light particles of St = 0.6 at $Re_\lambda$ =75 (Fig.~\ref{betaDependency4} (a)), and 180  (Fig.~\ref{betaDependency4} (b)). 
It clearly shows that the trends in the probability density functions are similar for both $Re_{\lambda}$.
The PDF of neutrally buoyant particles follows the $\Gamma$-distribution eq.\  
(\ref{Gamma function})  
quite well, reflecting that neutrally buoyant particles do not have any preferential concentration. In contrast, the PDFs of light and heavy particles clearly show a different behavior compared to the randomly distributed particles.
We observe that the probability of finding either small or large Vorono\"{\i} volumes is higher for both light and heavy particles. The two regions of small and large volumes can be used to identify clusters and voids. The strongest clustering is  observed for light particles, as the probability of finding small Vorono\"{\i} volumes is the highest.
 Owing to the density difference, light particles accumulate in vortex filaments due to centrifugal forces (\cite{Mazzitelli2003, Mazzitelli2004, Biferale2010}), while heavy particles concentrate in regions of intense strain (\cite{Bec2006}). Here, although the heavy particles show clustering, it is less compared to light particles. These results are consistent with the Minkowski analysis by \cite{Calzavarini2008c}.

\begin{figure}
\begin{center}
\includegraphics[width=1\textwidth]{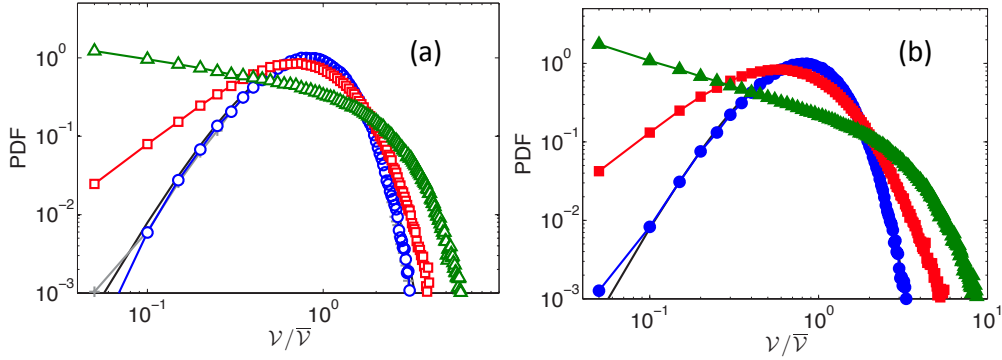}
\end{center}
\caption{ (Color online) The normalized Vorono\"{\i} volume PDFs for heavy (squares), neutrally buoyant (circles), and light particles (triangles) at St = 0.6 from DNS at (a) $Re_{\lambda} = 75$, and (b) $Re_{\lambda} = 180$. The thick line shows the $\Gamma$-distribution  eq.\ (\ref{Gamma function}) for randomly distributed particles (\cite{Ferenc2007}), the PDF of the neutrally buoyant particles agrees well with the randomly distributed particles ($+$). Both heavy and light particles show clustering, however light particles show the maximum clustering.} 
\label{betaDependency4}
\end{figure}

\subsection{Stokes number effect}

In this section, we study the effect of St on the clustering behavior for the three types of particles. We study the clustering behavior of the particles by examining the deviations of their Vorono\"{\i} volume PDFs from the $\Gamma-$distribution.

\begin{figure}
\begin{center}
\includegraphics[width=1\textwidth]{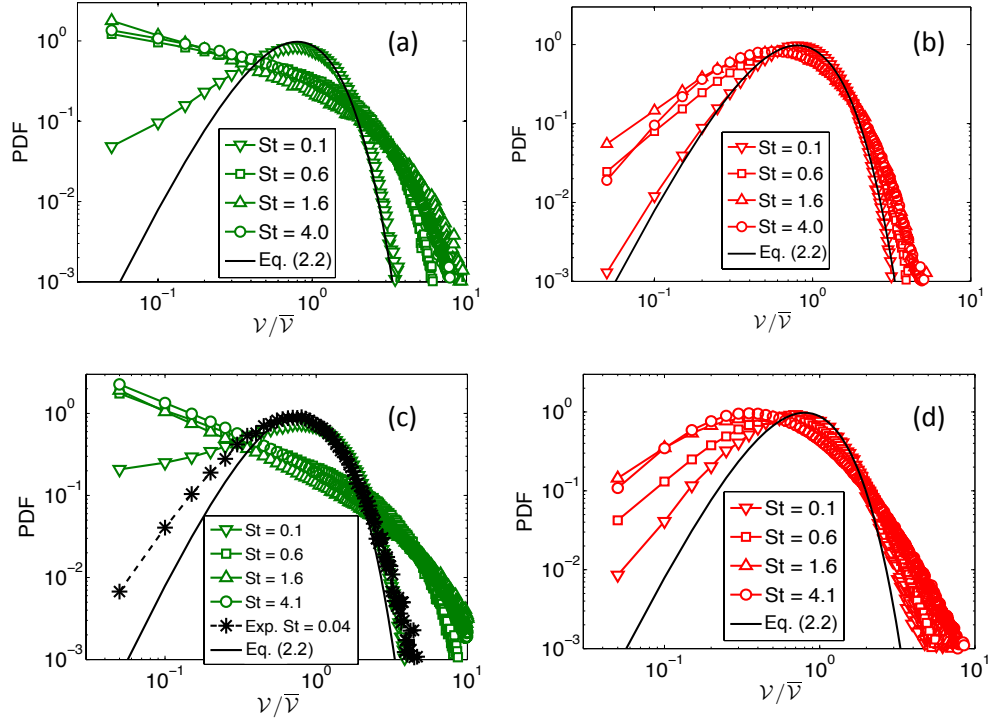}
\end{center}
\caption{ (Color online) The normalized Vorono\"{\i} volume PDFs  for different St ranging from 0.1 to 4 in the numerics for (a) light particles $\beta$=3 at $Re_{\lambda} = 75$,  (b) light particles $\beta$=3 at $Re_{\lambda} = 180$, (c) heavy particles $\beta$=0 at $Re_{\lambda} = 75$,  and (d) heavy particles $\beta$=0 at $Re_{\lambda} = 180$ .
The stars in (b)  correspond to the  experimental result with St = 0.04 $\pm$ 0.02 at $Re_{\lambda} = 162$.
}
\label{Stokes_effect}
\end{figure}

Figure~\ref{Stokes_effect} shows PDFs of light ($\beta$=3) and heavy ($\beta$=0) particles for different St at $Re_\lambda$ of 75 and 180. Firstly, we discuss the clustering of light particles as shown in Fig.~\ref{Stokes_effect} for $Re_\lambda$ of (a) 75 and (b) 180. 
Both Reynolds numbers give a similar trend with increasing St. When St increases, the probability of finding clusters and voids increases up to a value of St = 1.6, after which the dependence becomes weaker for both $Re_\lambda$. We note that the experimental result, shown with stars in Fig.~\ref{Stokes_effect} (b), for micro-bubbles with St =  0.04 $\pm$ 0.02 agrees reasonably well with the trend of the numerical data for 
light particles. In any case,  for these small Stokes numbers, the PDF of the Vorono\"{\i} volumes is still qualitatively similar to that of tracers. Another important feature of the light particle PDF is that the highest probability occurs at the smallest volume and decreases monotonically with increasing volume for St in the range of 0.6 to 4. As studied by \cite{Calzavarini2008c}, bubbles in this range of St tend to get trapped in vortex filaments, leaving void regions. Thus, most of the bubbles are concentrated in small regions and there are few bubbles outside these small regions.

In general, the clustering of heavy particles is weaker as compared to that of light particles. For heavy particles, as shown in Fig.~\ref{Stokes_effect} (c,d), as St increases, the probability of finding clusters and voids increases up to a value of St = 1.6, then the St dependence changes for different $Re_\lambda$ and is discussed below.

\begin{figure}
\begin{center}
\includegraphics[width=1\textwidth]{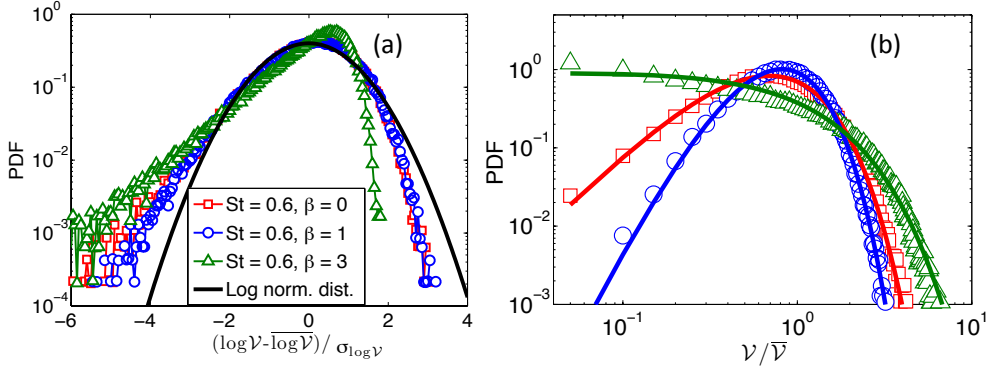}
\end{center}
\caption{ (Color online) The comparison between log-normal and $\Gamma-$distribution of Eq.~\ref{eq:fitting} fitting for the PDF of the 3D Vorono\"{\i} volumes. Open symbols represent heavy (squares), neutrally buoyant (circles), and light particles (triangles) at St = 0.6 from DNS at $Re_{\lambda} = 75$; the lines represent (a) log-normal, and (b) $\Gamma-$distribution.} 
\label{PDFfit}
\end{figure}

\cite{Monchaux2010} found that the Vorono\"{\i} area statistics of heavy particles can be well fitted by a log-normal distribution. For comparison, figure~\ref{PDFfit} (a) shows that the log-normal fitting for the PDFs of the present 3D Vorono\"{\i} volumes for the three different type of particles. It is clear that the Vorono\"{\i} volume statistics cannot be characterized well by the log-normal function, even not for the neutrally buoyant particles. However, the PDFs for all type of the particles can be fitted very well by the $\Gamma-$distribution (see figure~\ref{PDFfit} (b)) with only one fitting parameter $\sigma$:
\begin{equation}
f(x) = \frac{1}{\sigma^{\frac{2}{\sigma^2}}\Gamma(\frac{1}{\sigma^2})}x^{\frac{1}{\sigma^2}-1}\exp^{-\frac{x}{\sigma^2}}
\label{eq:fitting}
\end{equation}
where, $\sigma$ is the standard deviation of the Vorono\"{\i} volumes. Hence $\sigma$ provides a proper statistical quantification of Vorono\"{\i} volumes. 

In order to quantify the clustering using a single number, we use the standard deviation $\sigma$ of the normalized Vorono\"{\i} volume distributions. In figure~\ref{normsigma2} (a), we plot $\sigma$ normalized by the standard deviation of the Vorono\"{\i} volumes for randomly distributed particles $\sigma_{\Gamma}$. The magnitude of the indicator $\sigma/\sigma_{\Gamma}$ distinguishes the behavior of light, neutrally buoyant, and heavy particles. A higher value of the indicator reflects stronger clustering for a given $Re_\lambda$. For neutrally buoyant particles there is no observed clustering, hence the indicator value is constant at 1. 
Heavy particles show clustering and the indicator value saturates at St $\approx$ 1-2 at $Re_\lambda = $75. However, the indicator value continuously increases with St at the higher Reynolds number of $Re_\lambda = $180, and the absolute value of the indicator $\sigma/\sigma_{\Gamma}$ is larger for higher $Re_\lambda$. This indicates that the clustering of heavy particles is stronger at higher $Re_\lambda$ for a given St. 
Fig.~\ref{normsigma2} (a) shows that the absolute value of the indicator $\sigma/\sigma_{\Gamma}$ for light particles is also larger for higher $Re_\lambda$, revealing a stronger clustering for light particles at higher $Re_\lambda$. 
The reason for the Reynolds number effect could be because of the changing range of length scales of the vortex filaments which affect the clustering. At higher $Re_{\lambda}$, there is a wider range of clustering length scales resulting in a Vorono\"{\i} volume distribution with a higher value of standard deviation.
The curves corresponding to light particles show the strongest clustering, with a peak at St $\approx$ 1 $-$ 2 for both $Re_\lambda$ = 75 and 180. 
This clustering result has a consistent trend with that of the Kaplan-Yorke analysis (\cite{Calzavarini2008c}).

We also add the data point for the standard deviation of the experimental Vorono\"{\i} volume 
PDF as shown in figure~\ref{normsigma2}. Although the mean value of the indicator $\sigma/\sigma_{\Gamma}$ for the experimental data is higher than those from the numerical simulations
of light point particles, there is a good agreement with numerical trend within the experimental errorbar. More experimental data at larger Stokes numbers,
i.e., larger bubbles, have to be taken to come to a final conclusion on this issue.

\begin{figure}
\begin{center}
\includegraphics[width=1\textwidth]{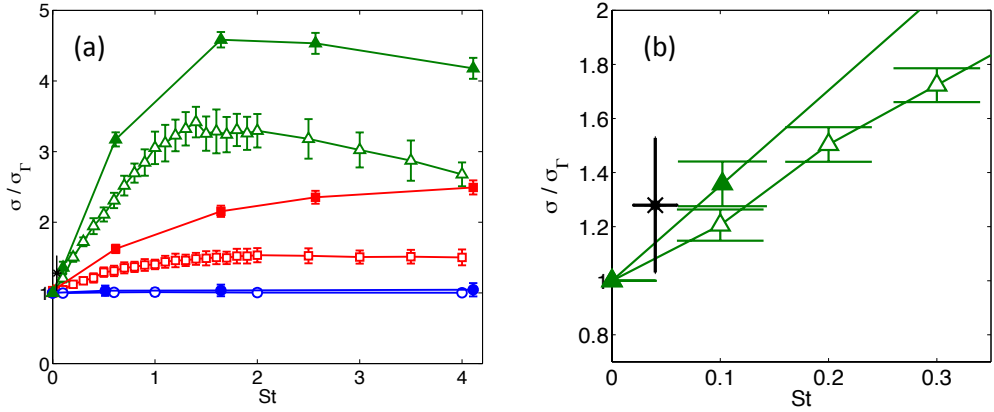}
\end{center}
\caption{ (Color online) (a) Normalized standard deviation $\sigma/\sigma_{\Gamma}$ (indicator) of the Vorono\"{\i} volume distributions versus St for the two different $Re_{\lambda}$ from DNS data. The symbols correspond to  heavy (squares), neutrally buoyant (circles), and light particles (triangles). Open and filled symbols represent data at $Re_{\lambda} = 75$ and $Re_{\lambda} = 180$, respectively. The value of the indicator for neutrally buoyant particles remains constant at 1, i.e. clustering is not observed, whereas light particles show the most clustering with a peak at St $\approx$ 1.5 for both $Re_{\lambda}$. The experimental result of micro-bubbles is plotted with the star. (b) An enlarged plot showing only the results for light particles.} 
\label{normsigma2}
\end{figure}

\subsection{Effects of the number of particles}

In principle, one can expect different behaviors depending on the number density of particles.
In the simulation dataset A, there are 10$^5$ particles available for one special case of $Re_\lambda$ = 75 and St = 0.6. Using this snapshot, we study the effects of the particle number on the value of the clustering indicator $\sigma/\sigma_{\Gamma}$. We subsample data from this snapshot by selecting the required number particles and computing the Vorono\"{\i} statistics. This subsampling procedure is randomized and then carried out at least 100 times for each case of particle number. Figure~\ref{fig:normsigma_space} shows the effect of varying the number of particles on the clustering indicator $\sigma/\sigma_{\Gamma}$ and the error bars represent the standard deviation of all the subsamples of a given number of particles.
In the present data set, the mean distances of particles are: 34.47$\eta$ for $N_{particles}$=10$^2$, 16$\eta$ for $N_{particles}$=10$^3$, 3.44$\eta$ for $N_{particles}$=10$^5$, which are all above 1$\eta$.
Hence, we are always studying situations where the mean particle distances are in the inertial range.

As shown in figure~\ref{fig:normsigma_space}, for light and heavy particles the value of the indicator increases as the number of the particles is increased. The evolution of the value of the indicator $\sigma/\sigma_{\Gamma}$ is steeper with increasing number of particles, and there seems to be no plateau region where the indicator value saturates. 
We do not understand the exact reason for this particle number dependence. One possible reason could be that the clusters have a complicated structure \cite[]{Calzavarini2008c}. 
However, for a given number of particles, the indicator does show a consistent trend: a stronger clustering for light particles, weaker clustering for heavy particles, and no clustering for neutrally buoyant particles. 
Moreover, the error of the indicator calculated at $N_{particle}$ = 1000 is less than 4 $\%$.
Therefore, at a fixed number of particles, the clustering indicator $\sigma/\sigma_{\Gamma}$ of the Vorono\"{\i} volume is robust.
In the analysis that follows, we use the data of $N_{particles}$ = 1000 for the simulation with the domain size of $N$ = 128$^3$ at $Re_\lambda$ = 75 (simulation A).

\begin{figure}
\begin{center}
\includegraphics[width=0.7\textwidth]{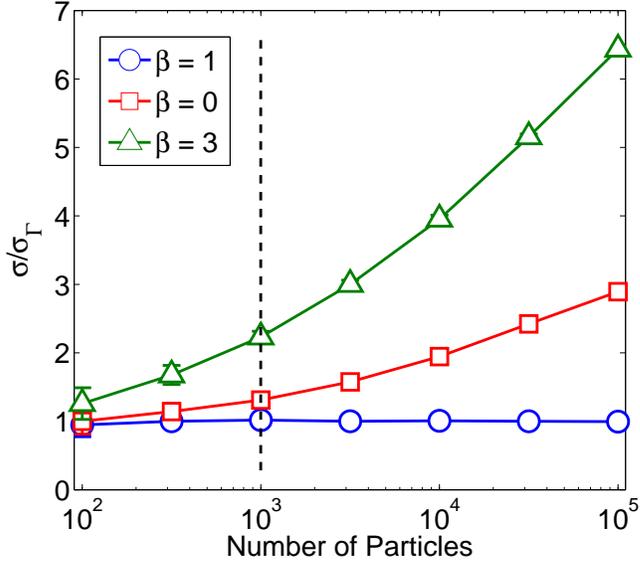}
\end{center}
\caption{ (Color online) Normalized standard deviation of the Vorono\"{\i} volume distributions as a function of number of particles taken from a snapshot case at $Re_{\lambda} = 75$. The dashed line shows the number of particles we used in the present work at $Re_{\lambda}$ = 75 and St = 0.6 (simulation A).} 
\label{fig:normsigma_space}
\end{figure}

\subsection{Relation between the volume of the Vorono\"{\i} cell and enstrophy}

We relate the Vorono\"{\i} volumes for the three different type of particles with turbulent flow quantities. A natural property  for this comparison would be the enstrophy $\Omega={\omega}^2/2$ (where $ \omega$ is vorticity). \cite{Benzi2009} have shown that different types of particles react sensitively to the local enstrophy at the particle position, reflecting their tendency to stay in regions with different vorticity contents. 
We thus calculate the joint PDF of Vorono\"{\i} volumes and enstrophy for three types of particles at a fixed St = 0.6 for $Re_\lambda =$ 75. 
For comparison, we also calculate the joint PDF for the case of neutrally buoyant particles with the smallest St available in the simulations (St = 0.1 and $\beta$ = 1). The statistical behavior of these particles is expected to be close to that of ideal fluid tracers (St = 0 and $\beta$ = 1). From now on, we refer to this case as the fluid tracer case. The Vorono\"{\i} volume and the enstrophy are normalized by the mean values ($\mathcal{V}_{tr}$ and $\Omega_{tr}$) of the fluid tracers. 
Figure~\ref{JPDFEnstVoro} shows the joint PDFs of the normalized Vorono\"{\i} volume ($\mathcal{V}/\mathcal{V}_{tr}$) and the normalized enstrophy ($\Omega/\Omega_{tr}$) for the different type of particles. 
The joint PDF for neutrally buoyant particles of St = 0.6, shown in figure~\ref{JPDFEnstVoro} (c), is very similar to that of fluid tracers shown in figure~\ref{JPDFEnstVoro} (a). We observe a clear difference in the joint PDF for heavy and light particles, as shown in figure~\ref{JPDFEnstVoro} (b, d). 
The coordinates corresponding to the peak of the joint PDF ($(\Omega/\Omega_{tr})_{jPDF}^{max}$, $(\mathcal{V}^p/\mathcal{V}_{tr})_{jPDF}^{max}$) is indicated by the crosses in the figure for each case. 
Compared to the tracer case, a slightly lower $(\mathcal{V}^p/\mathcal{V}_{tr})_{jPDF}^{max}$ and a lower $(\Omega/\Omega_{tr})_{jPDF}^{max}$ for heavy particles indicates more clustering at low enstrophy regions. The maximum value of the joint PDFs for the light particles is located at the region with a much higher enstrophy and a smaller Vorono\"{\i} volume. This shows that the light particles shows strong clustering at high enstrophy regions.

\begin{figure}
\begin{center}
\includegraphics[width=1\textwidth]{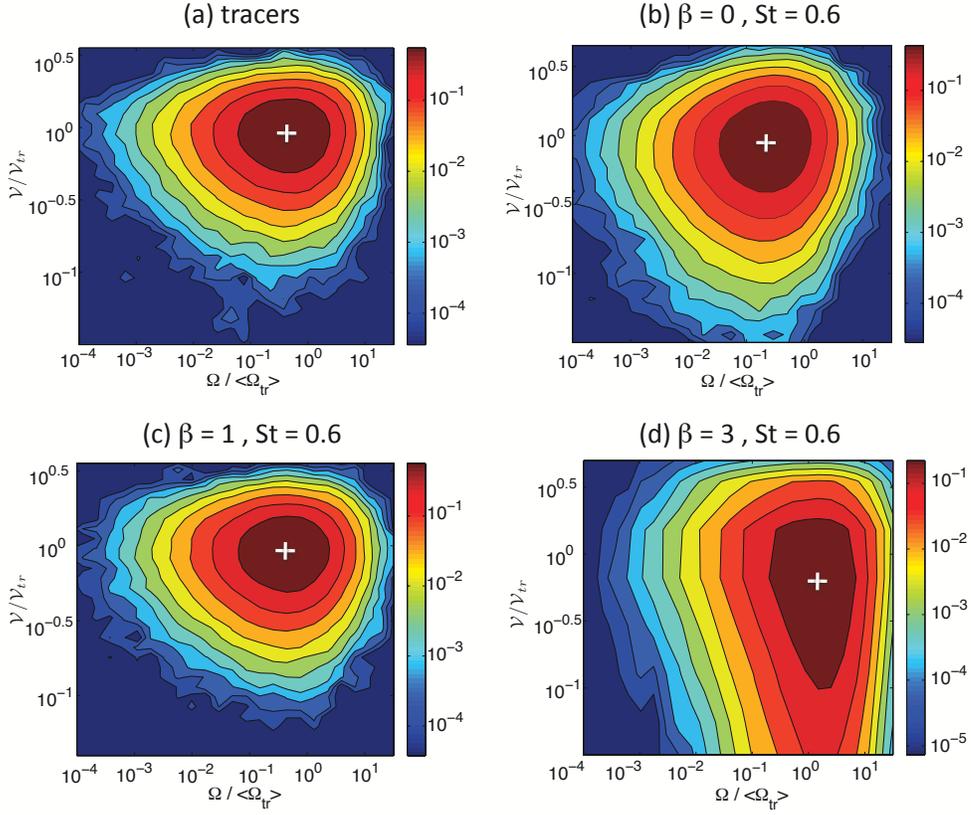}
\end{center}
\caption{ (Color online) Joint PDFs of normalized Vorono\"{\i} volumes and enstrophy for tracers and particles at St = 0.6 for $Re_\lambda =$ 75: (a) fluid tracers, (b) heavy particles, (c) neutrally buoyant particles, and (d) light particles. The cross indicates the location of the maximum probability (peak) for each case.}
\label{JPDFEnstVoro}
\end{figure}

\begin{figure}
\begin{center}
\includegraphics[width=1\textwidth]{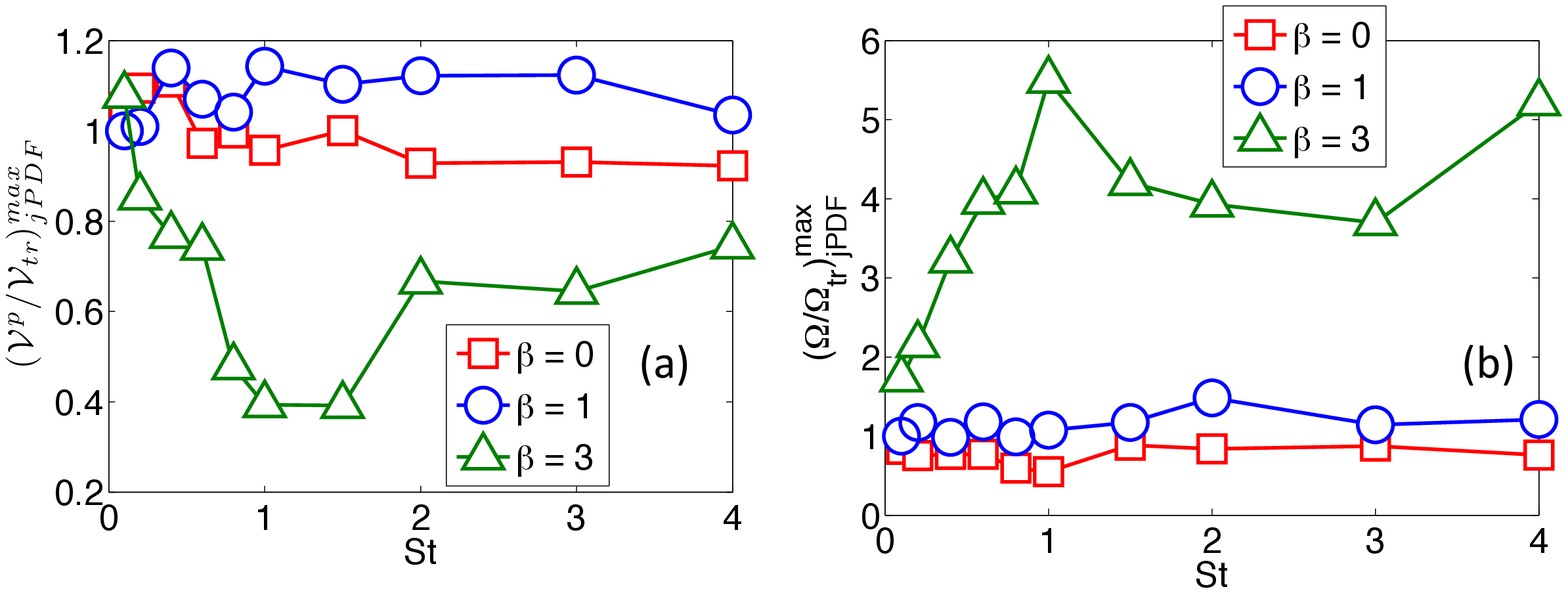}
\end{center}
\caption{ (Color online) The coordinates of the peak of the Joint PDFs of normalized Vorono\"{\i} volumes and enstrophy as a function of St: (a) $\mathcal{V}^p/\mathcal{V}_{tr}$ versus St, (b) $\Omega^p/\Omega_{tr}$ versus St.   }
\label{peak_st}
\end{figure}

The St dependence on the peak coordinates of the joint PDF ($(\Omega/\Omega_{tr})_{jPDF}^{max}$, $(\mathcal{V}^p/\mathcal{V}_{tr})_{jPDF}^{max}$) is plotted in figure~\ref{peak_st}. As shown in figure~\ref{peak_st} (a), the value of $(\mathcal{V}^p/\mathcal{V}_{tr})_{jPDF}^{max}$ for neutrally buoyant particles is nearly same with that of tracers at St from 0.1 to 4. The value of $(\mathcal{V}^p/\mathcal{V}_{tr})_{jPDF}^{max}$ for heavy particles is slightly smaller than unity for all St, indicating clustering. Figure~\ref{peak_st} (a) also shows that the clustering for light particles is stronger as evidenced by the much smaller $(\mathcal{V}^p/\mathcal{V}_{tr})_{jPDF}^{max}$ compared to those of neutrally buoyant and heavy particles at all St. The minimum value of $(\mathcal{V}^p/\mathcal{V}_{tr})_{jPDF}^{max}$ indicating strongest clustering for the light particles is located at St = 1 $-$ 2, which is in excellent agreement with the results obtained using the indicator $\sigma/\sigma_{\Gamma}$ (figure~\ref{normsigma2}). 
 The corresponding enstrophy at the peak ($(\Omega/\Omega_{tr})_{jPDF}^{max}$) of the joint PDF versus St for the different particles is shown in figure~\ref{peak_st} (b). The value of $(\Omega/\Omega_{tr})_{jPDF}^{max}$ for the heavy particles is smaller than unity, and it is much larger than unity for the light particles. This reflects the clustering of light particles in flow regions with very high enstrophy, whereas heavy particles cluster in low enstrophy regions for all St in the present study.

\subsection{Vorono\"{\i} Lagrangian autocorrelation}

Finally, we conduct a Lagrangian analysis on the Vorono\"{\i} volumes. For each type of particle we calculate the Lagrangian autocorrelation of its associated Vorono\"{\i} volume. Figure~\ref{AutoCorr} (a) shows a typical temporal evolution of Vorono\"{\i} volumes for the three types of particles at St = 0.6 and $Re_\lambda = $ 75. To compare the behavior of the three different particles, we choose particles with similar Vorono\"{\i} volume at the starting time and trace their time evolution. While the Vorono\"{\i} volumes of heavy and neutrally buoyant particles change frequently in time, it is clearly seen that light particles tend to have small values for longer times. This suggests that light particles are trapped in clustered regions for a long time and are suddenly ejected, as seen in figure~\ref{AutoCorr} (a) around $\tau/\tau_{\eta} \approx$ 95.

Figure~\ref{AutoCorr} (b) shows the autocorrelation function C$_{V}(\tau)$ for heavy, neutrally buoyant, and light particles at a fixed St = 0.6 and $Re_\lambda = $ 75. We define the decorrelation time $\tau_{V}$ as the time when the autocorrelation function has decreased to 1/2, i.e.,
 C$_{V}(\tau_V)$ = 1/2.
As shown in figure~\ref{AutoCorr} (b), the decorrelation time for light particles is around $\tau_{V} \sim 7 \tau_{\eta}$, whereas for heavy and neutrally buoyant particles 
decorrelation  already occurs 
around $4 \tau_{\eta}$. Thus the clustering of light particles lasts for a longer time as compared to heavy and neutrally buoyant particles.  
As shown by \cite{Calzavarini2008c}, light particles accumulate in filamentary structures and heavy ones tend to cluster outside these structures to form wall-like interconnected tunnels. 
These differences in the morphology of the clustered particles could be a possible reason for the light particles being clustered for a longer time as compared to heavy particles.

We also compare the autocorrelation time scale
of the Vorono\"{\i} volumes to that of the enstrophy shown in figure~\ref{AutoCorr}(c) for the same St and $Re_\lambda$.
 First, as expected, for neutrally buoyant particles, the Lagrangian decorrelation time for the Vorono\"{\i} volumes is comparable to that of the enstrophy ($\tau_{\Omega}$), 
i.e. $\tau_{\Omega} \sim \tau_{V}  \sim 4\tau_{\eta}$, because the neutrally buoyant particles do not cluster.
However, remarkably, for light particles  the decorrelation time of the Vorono\"{\i} volumes 
 is much larger,  $\tau_{V} \sim 7 \tau_{\eta}$, i.e., more than twice as large
as the autocorrelation time scale   $\tau_{\Omega} \sim 3 \tau_{\eta}$ of the enstrophy itself. 
For heavy particles, the Lagrangian decorrelation time of the Vorono\"{\i} volumes is around 
$\tau_{V} \sim 4 \tau_{\eta}$, which is also about two times that of enstrophy $\tau_{\Omega} \sim 2 \tau_{\eta}$.

We also study the St dependence of the decorrelation time scales of Vorono\"{\i} volume ($\tau_{V}$) and enstrophy ($\tau_{\Omega}$)  at $Re_\lambda  =$ 75 for heavy, neutrally buoyant, and light particles as shown in figure~\ref{AutoCorr_St} (a). We observe that $\tau_V$  for light particles is always larger than heavy and neutrally buoyant particles in the St range 0.1 to 4, with a peak around St unity.
This suggests that the light particles cluster for a longer time in the range of studied St. 

It is well known that flow regions of high enstrophy trap bubbles and regions with intense strain accumulate heavy particles. 
Figure~\ref{AutoCorr_St}(a) also shows that $\tau_{V}$ for both light and heavy particles is much larger than their decorrelation time of enstrophy $\tau_{\Omega}$ for all St from 0.1 to 4.
This is more clearly seen in figure~\ref{AutoCorr_St}(b) where the ratio $\tau_{V}/\tau_{\Omega}$ for both light and heavy particles is greater than unity for all St, while this ratio is always close to unity for the neutrally buoyant particles. 
Remarkably, this means that the life-times of the clustered bubbles and heavy particles are much longer than the life-time of the trapping flow structures themselves. 
The interpretation is that clustered particles are constrained in different regions of the flow and due to their inertia need time to reorganize themselves in the flow after sudden changes in flow conditions.
However, neutrally buoyant particles do not have this constraint and are distributed more evenly at any given time in the flow. 
Figure~\ref{AutoCorr_St}(b) shows that the ratio $\tau_{V}/\tau_{\Omega}$ for light particles has a weakly decreasing trend at St larger than unity. The ratio $\tau_{V}/\tau_{\Omega}$ for heavy particles monotonically increases with increasing St, and it is larger compared to light particles for St $>$ 0.5.

\begin{figure}
\begin{center}
\includegraphics[width=0.8\textwidth]{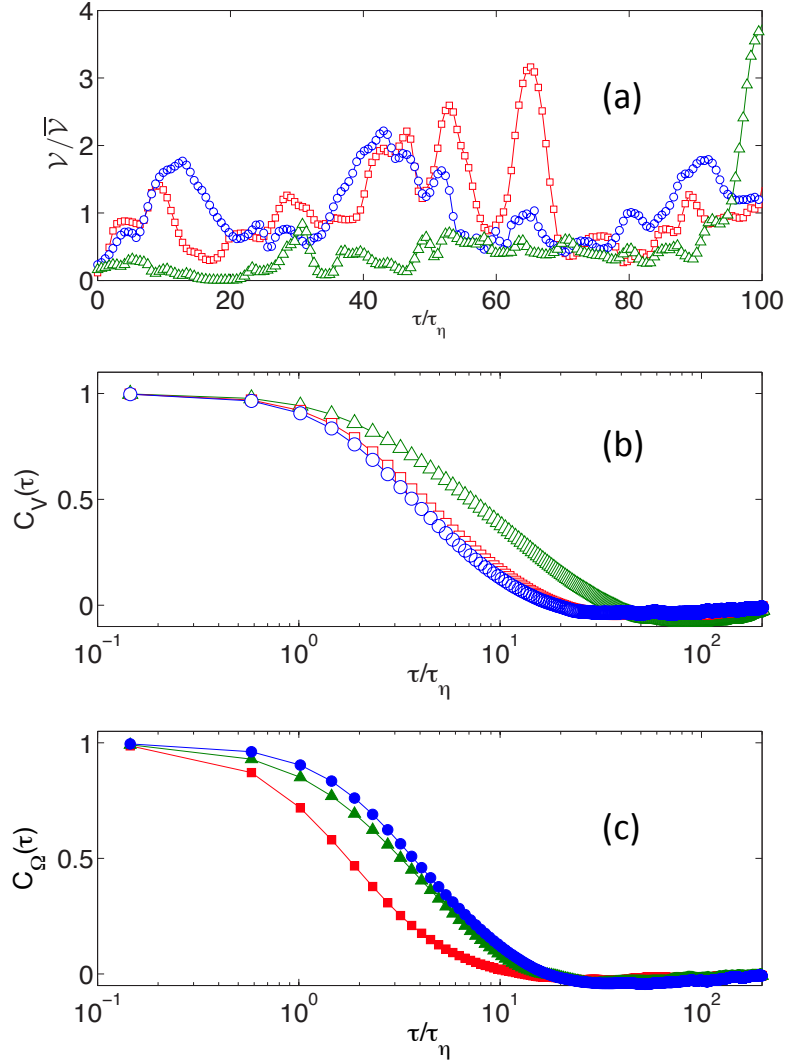}
\end{center}
\caption{ (Color online) Lagrangian Vorono\"{\i} analysis for heavy (squares), neutrally buoyant (circles), and light particles (triangles) at St = 0.6 and $Re_{\lambda} = 75$. (a) Temporal evolution of Vorono\"{\i} volumes.  (b) Temporal autocorrelation functions of Vorono\"{\i}  volumes. (c) Temporal autocorrelation functions of enstrophy.}
\label{AutoCorr}
\end{figure}

\begin{figure}
\begin{center}
\includegraphics[width=1\textwidth]{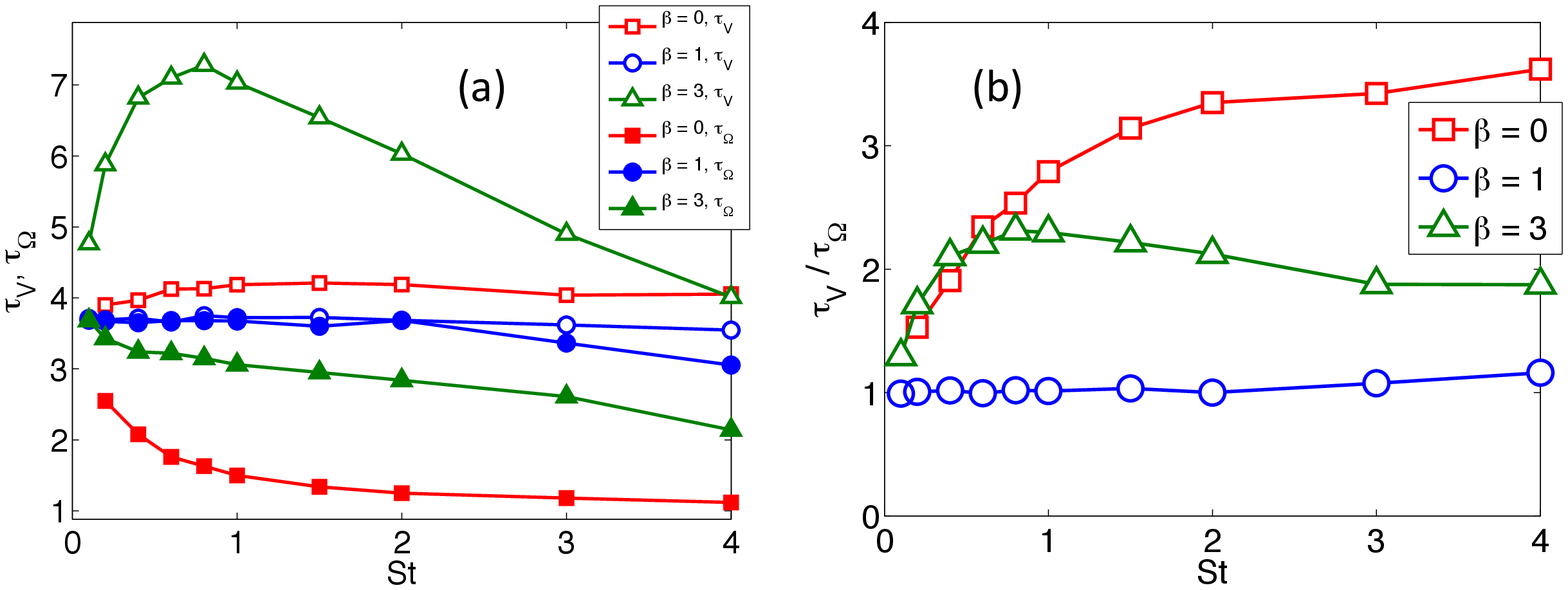}
\end{center}
\caption{ (Color online) (a) Decorrelation time of Vorono\"{\i} volume ($\tau_V$) and enstrophy ($\tau_{\Omega}$) as a function of St at  $Re_{\lambda} = 75$ for heavy (squares), neutrally buoyant (circles), and light particles (triangles). Open and filled symbols represent decorrelation time of Vorono\"{\i} volume and enstrophy, respectively. (b) The ratio of decorrelation times ($\tau_V$ / $\tau_{\Omega}$) as a function of St.}
\label{AutoCorr_St}
\end{figure}

\section{Conclusion} \label{sec:con}
We use three-dimensional Vorono\"{\i} analysis to study particle clustering in homogenous isotropic turbulence with both numerical data in the point particle limit and one experimental data set. The analysis is applied to inertial particles (light, neutrally buoyant, and heavy) of different density ratios $\beta$, St ranging from 0.1 to 4 and two different Taylor-Reynolds numbers ($Re_\lambda$ = 75 and 180). In the entire range of parameters covered, the Vorono\"{\i} volume PDFs of neutrally buoyant particles agree well with the $\Gamma$-distribution  for randomly distributed particles. At a fixed value of St, the PDFs of Vorono\"{\i} volumes of light and heavy particles show higher probability to have small and large Vorono\"{\i} volumes than randomly distributed particles, reflecting the clustering behavior. 
The standard deviation of normalized Vorono\"{\i} volumes $\sigma/\sigma_{\Gamma}$ is used as an indicator to quantify the clustering. 
Heavy particles show some clustering, and light particles have a much stronger clustering. Both heavy and light particles show a stronger clustering for a higher $Re_\lambda$. 
The maximum clustering for light particles is around St $\approx$ 1-2 for both Taylor-Reynolds numbers, and this maximum clustering range has a consistent trend with that of the Kaplan-Yorke analysis. 
We check the effect of number of particles on the value of the indicator and find that the clustering trend is robust for a given number of particles.

For one (small) Stokes number St = 0.04 $\pm$ 0.02
we have also extracted the  3D Vorono\"{\i} volume PDF  from experimental data.
Though the PDF fits into the general trend -- at these small Stokes numbers the PDF nearly follows
a $\Gamma$-distribution -- a quantitative analysis shows that the experimental PDF of 3D Vorono\"{\i}
 volumes is slightly broader than what is obtained from point-particle simulations. More experiments
with larger Stokes numbers will have to be done to judge whether this is a limitation of the
point-particle approach, a consequence of the neglectance of two-way and four-way coupling in 
the numerics, or whether the experimental data are not precise enough. 
From our point of view, the Vorono\"{\i} analysis is an excellent means to quantitatively compare clustering effects of particles 
in experimental and numerical data sets. 

Finally, we show that the Vorono\"{\i} analysis can be
 connected to local flow properties like enstrophy. By comparing the joint PDFs of enstrophy and Vorono\"{\i} volumes and their Lagrangian autocorrelation, the clustering behavior of heavy, neutrally buoyant, and light particles can further be distinguished. 
 It is found that the light particles strongly cluster in flow regions with very high enstrophy, whereas heavy particles weakly cluster in low enstrophy regions for all St in the present study.
From the Lagrangian autocorrelation of Vorono\"{\i} volumes we conclude that the clustering of light particles lasts much longer than that of heavy or neutrally buoyant particles. And due to inertial 
effects owing to the density difference from the carrying fluid, light and heavy particles
remain clustered for a much longer time than the flow structures themselves.

\section*{Acknowledgments}
We would like to acknowledge the support from COST Action MP0806: \emph{Particles in turbulence}. We acknowledge Mickael Bourgoin, Romain Monchaux, Sander G. Huisman, Ceyda Sanli, Devaraj van der Meer, and Federico Toschi for useful discussions. J.M.M. acknowledges the foundation for Fundamental Research of Matter (FOM) for the funding within the Industrial Partnership Programme: \emph{Fundamentals of heterogeneous bubbly flows}.


\begin{thebibliography}{22}
\expandafter\ifx\csname natexlab\endcsname\relax\def\natexlab#1{#1}\fi

\bibitem[Aliseda {\em et~al.\/}(2002)Aliseda, Cartellier, Hainaus \&
  Lasheras]{Aliseda2002}
{\sc Aliseda, A., Cartellier, A., Hainaux, F. \& Lasheras, J.C.} 2002 Effect of
  preferential concentration on the settling velocity of heavy particles in
  homogeneous isotropic turbulence. {\em J. Fluid Mech.\/} {\bf 468}, 77--105.

\bibitem[Bec {\em et~al.\/}(2006)Bec, Biferale, Boffetta, Celani, Cencini,
  Lanotte, Musacchio \& Toschi]{Bec2006}
{\sc Bec, J., Biferale, L., Boffetta, G., Celani, A., Cencini, M., Lanotte, A.,
  Musacchio, S. \& Toschi, F.} 2006 Acceleration statistics of heavy particles
  in turbulence. {\em J. Fluid Mech.\/} {\bf 550}, 349--358.

\bibitem[Biferale {\em et~al.\/}(2010)Biferale, Scagliarini \& Toschi]{Biferale2010}
{\sc Biferale L., Scagliarini, A. \& Toschi, F.} 2010 On the measurement of vortex filament lifetime statistics in turbulence. {\em Phys. Fluids\/} {\bf 22}, 065101


\bibitem[Benzi {\em et~al.\/}(2009)Benzi, Biferale, Calzavarini, Lohse \&
  Toschi]{Benzi2009}
{\sc Benzi, R., Biferale, L., Calzavarini, E., Lohse, D. \& Toschi, F.} 2009
  Velocity-gradient statistics along particle trajectories in turbulent flows:
  The refined similarity hypothesis in the Lagrangian frame. {\em Phys. Rev.
  E\/} {\bf 80}~(6), 066318.

\bibitem[Bodenschatz {\em et~al.\/}(2010)Bodenschatz, Malinowski, Shaw \&
  Stratmann]{Bodenschatz2010}
{\sc Bodenschatz, E., Malinowski, S.~P., Shaw, R.~A. \& Stratmann, F.} 2010 Can
  we understand clouds without turbulence? {\em Science\/} {\bf 327}, 970--971.

\bibitem[Calzavarini {\em et~al.\/}(2008{\natexlab{{\em a\/}}})Calzavarini,
  Cencini, Lohse \& Toschi]{Calzavarini2008a}
{\sc Calzavarini, E., Cencini, M., Lohse, D. \& Toschi, F.} 2008{\natexlab{{\em
  a\/}}} Quantifying turbulence-induced segregation of inertial particles. {\em
  Phys. Rev. Lett.\/} {\bf 101}, 084504.

\bibitem[Calzavarini {\em et~al.\/}(2008{\natexlab{{\em b\/}}})Calzavarini,
  Berg, Toschi \& Lohse]{Calzavarini2008b}
{\sc Calzavarini, E., van den Berg, T.H., Toschi, F. \& Lohse, D.} 2008{\natexlab{{\em
  b\/}}} Quantifying microbubble clustering in turbulent flow from single-point measurements
 {\em
  Phys. Fluids\/} {\bf 20}, 040702.

\bibitem[Calzavarini {\em et~al.\/}(2008{\natexlab{{\em c\/}}})Calzavarini,
  Kerscher, Lohse \& Toschi]{Calzavarini2008c}
{\sc Calzavarini, E., Kerscher, M., Lohse, D. \& Toschi, F.}
  2008{\natexlab{{\em c\/}}} Dimensionality and morphology of particle and
  bubble clusters in turbulent flow. {\em J. Fluid Mech.\/} {\bf 607}, 13--24.

\bibitem[Chen {\em et~al.\/}(2006)Chen, Goto \& Vassilicos]{Chen2006}
{\sc Chen, L., Goto, S. \& Vassilicos, J.C.} 2006 Turbulent clustering of
  stagnation points and inertial particles. {\em J. of Fluid Mech.\/}
  {\bf 553}, 143--154.

\bibitem[Ferenc \& N\'eda(2007)]{Ferenc2007}
{\sc Ferenc, J.S. \& N\'eda, Z.} 2007 On the size distribution of Poisson Vorono\"{\i} cells. {\em Physica A\/} {\bf 385}, 518--526.

\bibitem[Fessler {\em et~al.\/}(1994)Fessler, Kulick \& Eaton]{Fessler1994}
{\sc Fessler, J.R., Kulick, J.D. \& Eaton, J.K.} 1994 Preferential
  concentration of heavy particles in a turbulent channel flow. {\em Phys.
  Fluids\/} {\bf 6}, 3742--3749.

\bibitem[IJzermans {\em et~al.\/}(2009{\natexlab{{\em \/}}}) IJzermans, R.H.A., Reeks, M.W., Meneguz, E., Picciotto, M., \& Soldati, A.]{IJzermans2009}
{\sc IJzermans, R.H.A., Reeks, M.W., Meneguz, E., Picciotto, M., \& Soldati, A.}
  2009{\natexlab{{\em \/}}} Measuring segregation of inertial particles in turbulence by a full Lagrangian approach {\em Phys. Rev. E\/} {\bf80}, 015302(R)

\bibitem[Kerscher {\em et~al.\/}(2001)Kerscher, Mecke, Schmalzing, Beisbart,
  Buchert \& Wagner]{Kerscher2001}
{\sc Kerscher, M., Mecke, K., Schmalzing, J., Beisbart, C., Buchert, T. \&
  Wagner, H.} 2001 Morphological fluctuations of large-scale structure: the
  pscz survey. {\em Astron. Astrophys.\/} {\bf 373}, 1--11.

\bibitem[Martinez-Mercado {\em et~al.\/}(2010)Martinez-Mercado, Chehata-Gomez,
  {v}an Gils, Sun \& Lohse]{Martinez2010}
{\sc Martinez-Mercado, J., Chehata-Gomez, D., {v}an Gils, D. P.~M., Sun, C. \&
  Lohse, D.} 2010 On bubble clustering and energy spectra in pseudo-turbulence.
  {\em J. Fluid Mech.\/} {\bf 650}, 287--306.

\bibitem[Martinez-Mercado {\em et~al.\/}(2011)Martinez-Mercado, Prakash,
  Tagawa, Sun \& Lohse]{Martinez2011}
{\sc Martinez-Mercado, J., Prakash, V.~N., Tagawa, Y., Sun, C. \& Lohse, D.}
  2011 Under preparation .

\bibitem[Maxey \& Riley(1983)]{Maxey1983}
{\sc Maxey, M.R. \& Riley, J.J.} 1983 Equation of motion for a small rigid
  sphere in a nonuniform flow. {\em Phys. Fluids\/} {\bf 26}, 883--889.

\bibitem[Mazzitelli \& Lohse(2004)]{Mazzitelli2004}
{\sc Mazzitelli, I.M. \& Lohse, D.} 2004 Lagrangian statistics for fluid
  particles and bubbles in turbulence. {\em New J. Phys.\/} {\bf 6}, 203.

\bibitem[Mazzitelli {\em et~al.\/}(2003)Mazzitelli, Lohse \&
  Toschi]{Mazzitelli2003}
{\sc Mazzitelli, I.M., Lohse, D. \& Toschi, F.} 2003 On the relevane of the
  lift force in bubbly turbulence. {\em J. Fluid Mech.\/} {\bf 488}, 283--313.

\bibitem[Monchaux {\em et~al.\/}(2010)Monchaux, Bourgoin \&
  Cartellier]{Monchaux2010}
{\sc Monchaux, R., Bourgoin, M. \& Cartellier, A.} 2010 Preferential
  concentration of heavy particles: A {V}orono\"{\i} analysis. {\em Phys. Fluids\/}
  {\bf 22}, 103304.

\bibitem[Okabe {\em et~al.\/}(2000)Okabe, Boots, Sugihara \&
  S.N.]{SpatialTesselation}
{\sc Okabe, A., Boots, B., Sugihara, K. \& S.N., Chiu} 2000 Spatial
  tesselations. {\em John Wiley \& Sons Ltd.\/} .

\bibitem[Pratsinis {\em et~al.\/}(1996)Pratsinis, \& Vemury] {Pratsinis1996}
{\sc Pratsinis S.E. \& Vemury, S.} 1996 Particle formation in gases: a review. 
{\em Powder Technology\/} {\bf 88}, 267--273.

\bibitem[Saw {\em et~al.\/}(2008)Saw, Shaw, Ayyalasomayajula, Chuang \&
  Gylfason]{Saw2008}
{\sc Saw, E.W., Shaw, R.A., Ayyalasomayajula, S., Chuang, P.Y. \& Gylfason, A.}
  2008 Inertial clustering of particles in high-{R}eynolds-number turbulence.
  {\em Phys. Rev. Lett.\/} {\bf 100}, 214501.

\bibitem[Schmitt \& Seuront(2008)]{Schmitt2008}
{\sc Schmitt, F.G. \& Seuront, L.} 2008 Intermittent turbulence and copepod
  dynamics: increase in encounter rates through preferential concentration.
  {\em Journal of Marine Systems\/} {\bf 70}, 263--272.

\bibitem[Toschi \& Bodenschatz(2009)]{Toschi2009a}
{\sc Toschi, F. \& Bodenschatz, E.} 2009 Lagrangian properties of particles in
  turbulence. {\em Ann. Rev. Fluid Mech.\/} {\bf 41}, 375--404.

\bibitem[Toschi {\em et~al.\/}(2009{\natexlab{{\em \/}}})Toschi, F., Biferale, L., Calzavarini, E., Scagliarini, A., \& Leveque, E.]{Toschi2009b}
{\sc Toschi, F., Biferale, L., Calzavarini, E., Scagliarini, A., \& Leveque, E.}
  2009{\natexlab{{\em \/}}} Lagrangian modeling and properties of particles with inertia. {\em Advances in Turbulence, XII, Proceedings of the 12th European Turbulence Conference (ETC-12), Marburg (D)\/} {\bf}, Springer Proceedings in Physics.

\bibitem[{van de}~Weygaert \& Icke(1989)]{Weygaert1989}
{\sc {van de}~Weygaert, R. \& Icke, V.} 1989 Fragmenting the universe.
  ii-Vorono\"{\i} vertices as abell clusters. {\em Astron. Astrophys.\/} {\bf 213},
  1--9.



\end{thebibliography}

\end{document}